# Highly bright photon-pair generation in Doppler-broadened ladder-type atomic system

Yoon-Seok Lee,[1] Sang Min Lee,[1,2] Heonoh Kim,[1] and Han Seb Moon,[1,*]

[1]*Department of Physics, Pusan National University, Geumjeong-Gu, Busan 46241, South Korea*
[2] *Currently with Korea Research Institute of Standards and Science, Daejeon 34113, South Korea*
*\*hsmoon@pusan.ac.kr*

**Abstract:** We report a bright photon-pair source with a coincidence counting rate per input power (cps/mW) of tens of thousands, obtained via spontaneous four-wave mixing from a Doppler-broadened atomic ensemble of the $5S_{1/2}$–$5P_{3/2}$–$5D_{5/2}$ transition of $^{87}$Rb. The photon-pair generation rate is enhanced by the two-photon coherence contributions from almost all the atomic velocity groups in the Doppler-broadened ladder-type atomic system. We obtained the violation of the Cauchy-Schwarz inequality by a factor of $2370 \pm 150$. We believe that our scheme for highly bright paired photons is important as a useful quantum light source for quantum entanglement swapping between completely autonomous sources.

## 1. Introduction

A bright single-photon source is important for studies on fundamental quantum optics and quantum communication including quantum key distribution and quantum teleportation [1]. However, it is difficult to implement an on-demand single-photon source. As is well-known, correlated photon pairs have been regarded as heralded single photons [2] for the realization of quantum networks [3,4]. The most popular method for the generation of correlated photon pairs is the spontaneous parametric down-conversion (SPDC) process in $\chi^{(2)}$ nonlinear crystals [5–7]. SPDC sources are highly bright photon pairs, but their coherence time is short. Recently, the generation of correlated photon pairs has been intensively studied using the spontaneous four-wave mixing (SFWM) process in various atomic systems [8–16]. SFWM sources have been demonstrated in cold [8–13] and Doppler-broadened [14–16] atomic media. The coherence time of SFWM sources in an atomic system is greater than that of SPDC sources, but the photon-pair generation rate of SFWM sources is low.

SFWM sources in a Doppler-broadened atomic ensemble have been achieved using a warm atomic vapor [14–16]. Although the SFWM sources in a Doppler-broadened atomic ensemble is simpler and more robust than those in a cold atomic ensemble, the performance of the SFWM source in the Doppler-broadened atomic ensemble is not sufficient for the desired brightness and bandwidth. Particularly, in the previous studies on a Doppler-broadened ladder-type atomic system [15,16], the broad range of velocity groups in the Doppler-broadened atomic ensemble did not entirely participate in the photon-pair generation. This is why the two-photon resonant condition of atoms interacting with two laser fields is not Doppler-free when the wavelength difference between the two laser fields is large. Two-photon coherence is essential for the enhancement of the third-order nonlinearity of the atomic medium. Therefore, the maximization of the two-photon coherence effect is very important for the generation of correlated photon pairs in a Doppler-broadened atomic ensemble.

In this paper, we experimentally demonstrate a highly bright photon-pair source via SFWM in the Doppler-broadened ladder-type atomic system of the $5S_{1/2}$–$5P_{3/2}$–$5D_{5/2}$ transition of $^{87}$Rb atoms. With high brightness comparable to that of SPDC and a long coherence time relative to the detection-time jitter, the full statistical properties in the time domain are directly measured via the Hanbury Brown-Twiss (HBT) experiment [17]. The photon pairs generated from the

Doppler-broadened atomic ensemble are characterized, considering the coincidence count rate as a function of the pump powers, the cross-correlation function, the auto-correlation function, and the second-order correlation function of the heralded single photon.

## 2. Experimental setup for photon-pair generation

The ladder-type atomic system used in this work is the $5S_{1/2}$–$5P_{3/2}$–$5D_{5/2}$ transition of $^{87}$Rb, as schematically shown in Fig. 1(a). The ladder-type atomic system consists of a ground state ($5S_{1/2}$), an intermediate state ($5P_{3/2}$), and an excited state ($5D_{5/2}$). The pump ($\Omega_p$) field interacts with the $5S_{1/2}$–$5P_{3/2}$ transition, and the coupling field ($\Omega_C$) interacts with the $5P_{3/2}$–$5D_{5/2}$ transition. The symbols $\delta_P$ and $\delta_C$ denote the detuning frequencies of $\Omega_p$ and $\Omega_C$, respectively. The signal and idler photons can be generated in the $5S_{1/2}$–$5P_{3/2}$–$5D_{5/2}$ transition under the phase-matching condition of the contributing fields. The natural linewidths of the $5P_{3/2}$ and $5D_{5/2}$ states are 6.065 MHz and 0.6673 MHz, respectively [18]. Decay from the $5D_{5/2}$ state to the $5S_{1/2}$ state via the $5P_{3/2}$ and $6P_{3/2}$ states is possible.

The experimental setup is shown in Fig. 1. The fields $\Omega_p$ and $\Omega_C$ were counter-propagated through a 12.5-mm-long vapor cell containing the $^{87}$Rb isotope and spatially overlapped completely with the same $1/e^2$ beam diameters of 1.2 mm. The intensities of both fields were adjusted using a half-wave plate and polarizing beam splitter, and the temperature of the vapor cell was set to 52 °C. The vapor cell was housed in a three-layer µ-metal chamber to reduce the external magnetic field. The signal and idler photons were generated via the SFWM process in the phase-matched direction and collected into two single-mode fibers (SMF).

In order to remove the uncorrelated fluorescence, we used interference filters and etalon filters. The interference filters have 3-nm bandwidth and 95% transmittance, and the etalon filters have a full width at half maximum (FWHM) linewidth of 950 MHz and 85% peak transmission. After passing through the filters, the signal and idler photons were coupled into multi-mode fibers. The estimated total efficiency for obtaining the spatially single-mode and noise-suppressed photon pairs was approximately 50%. The photons were detected by silicon avalanche photo-detectors (APD, PerkinElmer SPCM-AQRH-13HC). The statistical properties of the photon-pair source were investigated using the HBT experiment, which included a fiber beam splitter and time-correlated single-photon counter (TCSPC) in the start-stop mode with a 4-ps time resolution.

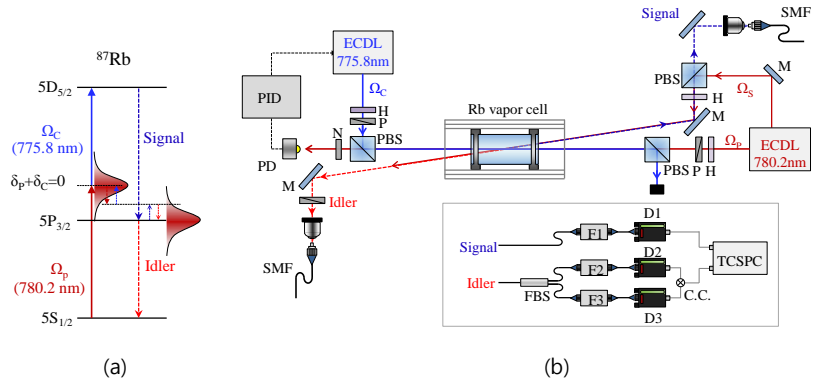

Fig. 1 (a) Cascade emission of signal and idler photons via spontaneous four-wave mixing in Doppler-broadened three-level ladder-type atomic system interacting with pump ($\Omega_p$) and coupling ($\Omega_C$) fields in the $5S_{1/2}$–$5P_{3/2}$–$5D_{5/2}$ transition of $^{87}$Rb atoms. (b) Experimental schematic for photon-pair generation in the atomic vapor cell of $^{87}$Rb with 780.2- and 776.8-nm ECDL for $\Omega_p$ and $\Omega_C$ fields, respectively (P: polarizer; H: half-wave plate; PBS: polarizing beam splitter; PD: photodiode; N: neutral density filter; M: mirror; F1-F3: interference filter; SMF: single mode fiber; FBS: fiber beam splitter; D1-–D3: single-photon detectors (SPDs); TCSPC: time-correlated single-photon counting module).

To conduct the experiment for photon-pair generation under the two-photon resonant condition, we observed the transmittance spectra of $\Omega_p$, as shown in Fig. 2. A narrow two-photon absorption (TPA) signal is clearly apparent, which implies that the two-photon coherence was strongly generated in the $5S_{1/2}$ (F = 2)–$5P_{3/2}$ (F' = 3)–$5D_{5/2}$ (F" = 4) transition of the atomic ensemble [19]. To maximize the two-photon coherence and prevent the Rayleigh scattering due to $\Omega_p$ and $\Omega_C$, the frequencies of both fields were stabilized at the 810-MHz blue-detuned two-photon absorption spectrum out of the Doppler absorption profile. We can see that the magnified TPA spectrum is narrow and strong, as shown in the inset of Fig. 2, because of the Doppler-free two-photon coherence due to the small wavelength difference. Therefore, the $5S_{1/2}$ (F = 2)–$5P_{3/2}$ (F' = 3)–$5D_{5/2}$ (F" = 4) transition is an excellent configuration for photon-pair generation in a Doppler-broadened atomic system.

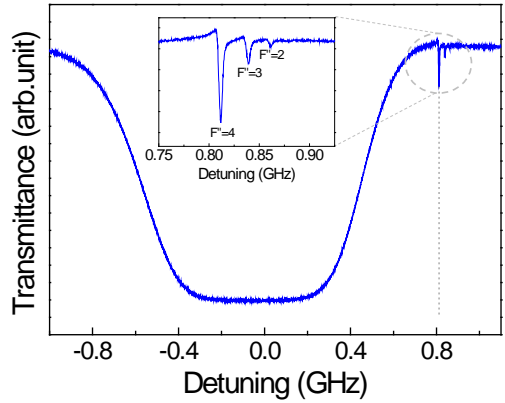

Fig. 2. Transmittance spectra of pump field as functions of frequency detuning of $5S_{1/2}$–$5P_{3/2}$–$5D_{5/2}$ transition of $^{87}$Rb atoms.

## 3. Experimental results and discussion

### 3.1 Photon counting rates

The photon counting rate is an important factor determining the properties of a photon-pair source. Firstly, we measured the counting rates of the signal photon of the $5P_{3/2}$ (F' = 3)–$5D_{5/2}$ (F" = 4) transition and the idler photon of the $5S_{1/2}$ (F = 2)–$5P_{3/2}$ (F' = 3) transition from the Doppler-broadened ladder-type atomic system shown in Fig. 1(a). Figure 3 shows the signal and idler single counting rates ($N_S$ and $N_I$, respectively) and the net coincidence counting rate ($N_C$) as functions of the pump power for different coupling powers (10, 20, 30, 40, and 48 mW) with a coincidence window of 4.1 ns, where all results were averaged over 10 measurements. Both single counting rates were corrected by considering a detector dead time of ~50 ns, whereas $N_C$ remained unchanged [20]. In the case of the coupling power of 48 mW, $N_C$ was 64,600 cps/mW with the pumping power of 1 mW, which is comparable to that of SPDC in a $\chi^{(2)}$ nonlinear crystal without other corrections such as filtering, coupling, and detection efficiency adjustments [7]. At the low coupling power of 10 mW, at which $N_C$ was 30,300 cps/mW excluding the effects of the dead time and saturation, both the single and coincidence counts increased as the pump power was increased to 1 mW. However, when the pump power was greater than 1 mW, a tendency towards the saturation of the photon counting results was observed. From the experimental results, the number of generated photon pairs $N_{pair}$ per input pump power was estimated to be 8.98(22) MHz/mW by using the equation $N_{pair} = (N_S \times N_I) / N_C$. This outstanding brightness was attributed to the collective enhancement of the emission rate due to the coherent contributions from almost all the velocity classes in the Doppler-broadened atomic ensemble. The heralding efficiencies $\eta_{S,I} = N_C / N_{S,I}$ for both signal and idler photons were calculated as 5.8(1)%, where filtering efficiency was not considered.

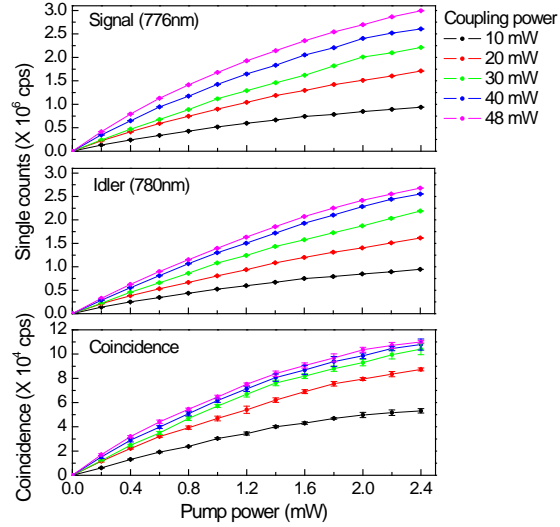

Fig. 3. Single counting rates for signal and idler photons and coincidence counting rate of photon pairs as functions of pump power for coupling powers of 10, 20, 30, 40, and 48 mW.

## 3.2 Second-order correlation functions

It is well known that a photon-pair source reliant on a spontaneous parametric process such as SPDC or SFWM exhibits strong temporal correlation between the signal and idler photons. . In our experiment, since the coherence time of generated photon pairs is a nanosecond order longer than the detection-timing uncertainty, we can directly resolve the temporal profile of the time correlation between the two photons, as well as the second-order auto-correlation function of the individual photons. To characterize the temporal statistical properties of the photons generated in this study, we measured the normalized second-order auto-correlation functions $g^{(2)}(\tau)$ for photons on each mode and both modes of the signal and idler, where $\tau$ is the time delay between two photon counting events measured using two single-photon detectors (SPDs), as shown in Fig. 4.

We performed the HBT experiment for individual signal and idler photons, as shown in Fig. 4(a). The collected signal and idler photons passed through a 50/50 fiber beam splitter and filters (F2 and F3) and were detected by two SPDs (D2 and D3). The histograms of the coincidence events between D2 and D3 as a function of $\tau$ show the auto-correlation functions $g_{SS}^{(2)}(\tau)$ and $g_{II}^{(2)}(\tau)$ of the signal and idler photons, respectively, and the statistical properties of the individual photons. Because of the parametric process, we expect that the photons in each mode exhibit the photon statistical property of thermal light. However, it is difficult to observe the bunching effect of photons in individual modes of the SPDC source because the coherence time is short compared with the response time of commercial SPDs. However, in our system, the relatively long coherence time of the paired photons allows the direct and clear observation of the bunching of individual photons, as shown in Fig. 4(a). For the signal and idler modes, the second-order auto-correlation values for $\tau = 0$ were measured as $g_{SS}^{(2)}(0)$ =1.74(9) and $g_{II}^{(2)}(0)$ =1.74(6), respectively. The red solid curves are Gaussian fits to the experimental results. The degree of bunching of the individual photons from the parametric process is related to the ratio of the coherence time to the detection jitter [21]. A full theoretical description incorporating the auto-correlation and multi-photon effect can be presented according to the quantum Langevin theory [22]. Note that the residual uncorrelated photons from the Rayleigh scattering and detection-time jitter are the main factors affecting the imperfect observation of the thermal nature.

As shown in Fig. 4(b), we obtained the cross-correlation function between the signal and idler photons by measuring the coincident detection histogram as a function of $\tau$. The normalized cross-correlation function $g_{SI}^{(2)}(\tau)$ for 1-mW pump power and 10-mW coupling power is shown in Fig. 4(b). Here, $g_{SI}^{(2)}(\tau)$ is defined as

$$g_{SI}^{(2)}(\tau) = \frac{G_{SI}^{(2)}(\tau)}{N_S N_I \Delta \tau T}, \qquad (1)$$

where $G_{SI}^{(2)}(\tau)$ is the second-order cross-correlation function for the paired photon generated from a Doppler-broadened atomic ensemble via SFWM. $G_{SI}^{(2)}(\tau)$ is the measured coincidence histogram. We averaged the histogram over $\Delta\tau = 300$ ps after measurement with a 4-ps time resolution for $T = 30$ s. The peak value $g_{SI}^{(2)}(0) = 84.70(1)$ exhibited a strong temporal correlation, which implies that the idler-photon emission was enhanced in the phase-matched direction upon heralding detection of the signal photon. The FWHM of $g_{SI}^{(2)}(\tau)$ was estimated to be 1.9(3) ns, corresponding to the inverse of the Doppler broadening of 540 MHz, which is significantly shorter than the excited-state lifetime of 26.24(4) ns in the $^{87}$Rb D$_2$ transition.

The second-order cross-correlation function for the paired photon generated from a Doppler-broadened atomic ensemble via SFWM is defined as [15]

$$G_{SI}^{(2)} = \left| \int \Psi_v(\tau) f(v) dv \right|^2, \qquad (2)$$

where $\Psi_v(\tau)$ is the two-photon amplitude for the SFWM process in a Doppler-broadened three-level ladder-type atomic system and $f(v)$ is the one-dimensional Maxwell-Boltzmann velocity distribution function.

As the entire broad range of velocity groups in the Doppler-broadened atomic ensemble can coherently contribute to photon-pair generation, $G_{SI}^{(2)}(\tau)$ is significantly enhanced via constructive interference between the two-photon amplitudes from different velocity groups. The red solid line in Fig. 4(b) is the fitted theoretical curve of Eq. (2), and the temporal shape of the $g_{SI}^{(2)}(\tau)$ histogram is in good agreement with the theoretical curve.

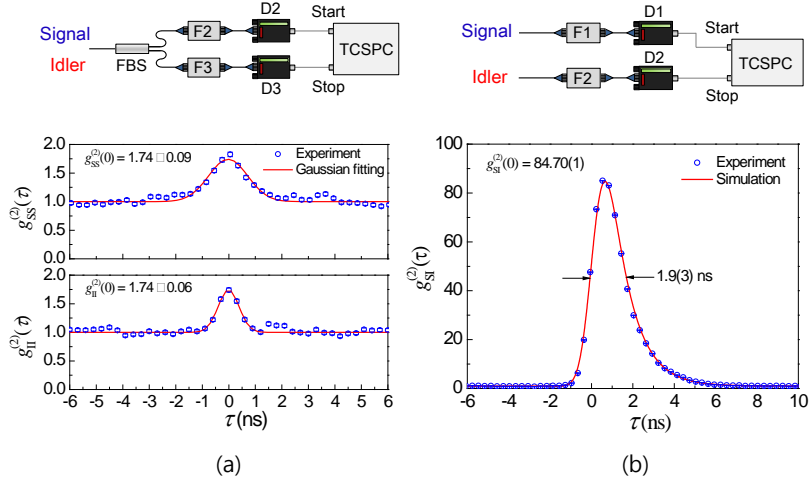

Fig. 4 Measurement of temporal statistical properties of generated signal and idler photons. (a) Normalized auto-correlation function for individual signal and idler photons measured in the Hanbury Brown-Twiss experiment; the red solid lines are Gaussian fitting curves. (b) Normalized temporal cross-correlation function between signal and idler photons; the red solid curve is the result calculated using Eq. (1).

Additionally, we estimated the Cauchy-Schwarz inequality from the normalized second-order correlation function of Fig. 4 in order to clarify that the photon pair has a temporal non-classical quantum correlation [23]. As is well known, the Cauchy-Schwarz inequality is described as

$$R = \frac{\left[g_{SI}^{(2)}(\tau)\right]^2}{g_{SS}^{(2)}(\tau)\, g_{II}^{(2)}(\tau)} \leq 1. \quad (3)$$

In this work, the $R$ value calculated using Eq. (3) for the $g^{(2)}(0)$ values of Fig. 5 at $\tau = 0$ violated the Cauchy-Schwarz inequality by a factor $R = 2370 \pm 150$, when the pump and coupling powers are 1 mW and 10 mW, respectively. The main factors causing the large $R$ value are the enhanced SFWM photon-pair flux achieved by exploiting all the velocity classes in the Doppler-broadened atomic medium and the suppressed single-emission fluorescence achieved using far-detuned pump and coupling fields with relatively weak pump power.

### 3.3 Photon statistics of the heralded single photons

We investigated the photon statistics for the heralded single photons by using the conditional HBT experiment of Fig. 5(a) [24]. When the coincidence event between D1 and D2 occurred, the second-order correlation function for the idler photon was measured by detecting the coincidence of D3 and the heralded signal photon. Figure 5(b) shows the three-fold coincidence histogram as a function of the time difference ($\tau$) between the coincidence event (D1 and D2) with a coincidence window of 3.3 ns, and the single-counting event (D3) with a 300-ps time bin. Here, taking advantage of the high brightness of our source, we set the 1-mW pump and 40-mW coupling powers, where $N_S$ and $N_I$ were 1,420,000 cps and 1,300,000 cps, respectively, and $N_C$ was 61,700 cps. Hence, the data-acquisition time was reduced to $T = 300$ s for the measurement of the three-fold coincidence histogram. The temporal shape of $g_{SII}^{(2)}(\tau)$ is similar to that of $g_{SI}^{(2)}(\tau)$; the peak value of $g_{SII}^{(2)}(\tau)$ was measured to be 3.88(7). Theoretically, the conditional auto-correlation function ($g_{SII}^{(2)}(\tau)$) has a maximum peak value of 4, which is the result of thermal bunching of the individual photons and the strong time correlation between the signal and idler photons. The difference from the experimental result and the theoretically expected value of 4 is due to the coincidence window, along with the factors that introduce imperfections in the observation of the bunching effect of the individual photons, i.e., the Rayleigh scattering and detection-time jitter.

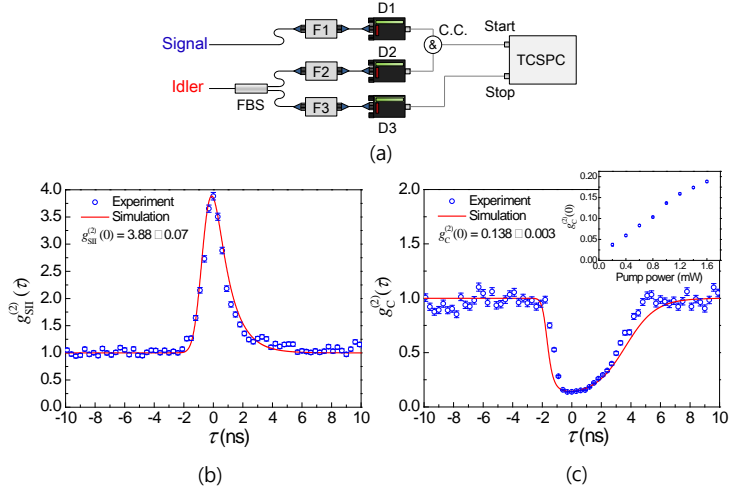

Fig. 5. (a) Schematic of the conditional HBT experimental setup for obtaining the second-order correlation function for the heralded single photon. (b) Temporal histogram measured in the conditional HBT experiment (the red solid curve is the numerical result). (c) Normalized second-order correlation function for heralded single photon in idler mode (the red solid curve is the numerical result). Inset: Value of $g_C^{(2)}(0)$ as a function of pump power with 40-mW coupling power.

To intuitively reveal the single-photon nature of our heralded single photon, we constructed the normalized standard second-order correlation function $g_C^{(2)}(\tau)$ of Fig. 5(c) from the results of Fig. 4 and 5(b). The normalized second-order correlation function for the heralded idler photons is

$$g_C^{(2)}(\tau) = \frac{\overline{G}_{SI_1I_2}^{(2)}(\tau) R(0)}{\overline{N}_{SI_1}^{(2)}(0) \overline{G}_{SI_2}^{(2)}(\tau)} . \tag{4}$$

Here, $\overline{G}_{SI_1I_2}^{(2)}(\tau)$ is the time-averaged temporal three-fold coincidence histogram over the 300-ps time bin between the two-fold coincidence counting event (D1 and D2) with a 3.3-ns coincidence window and the single-counting event (D3). $\overline{G}_{SI_2}^{(2)}(\tau)$ is the time-averaged temporal histogram of two-fold coincidence events (D1 and D3). $R(0)$ and $\overline{N}_{SI_1}^{(2)}(0)$ are experimentally measured from the single-counting rate of the signal photon and the coincidence counting rate between D1 and D2, respectively.

The $g_C^{(2)}(\tau)$ of Fig. 5(c) shows the anti-bunching feature of the heralded single photon. The minimum value of $g_C^{(2)}(\tau)$ was estimated to be 0.138(3) at $N_S = 1.42 \times 10^6$ cps, as shown in Fig. 5(c). The inset of Fig. 5(c) shows the pump-power dependence of $g_C^{(2)}(0)$ at 40-mW coupling power. As the multi-photon detection probability decreases with decreasing pump power, the $g_C^{(2)}(0)$ value decreases. When the pump power was reduced to 0.2 mW, we obtained $g_C^{(2)}(0) = 0.037(3)$.

## 4. Conclusion

We have demonstrated highly bright photon pairs at wavelengths of 780.2 nm and 775.8 nm via spontaneous four-wave mixing from a warm Rb atomic medium. The entire broad range of velocity groups in the Doppler-broadened atomic ensemble contributed to generate strong two-photon coherence using the counter-propagating geometry in the $5S_{1/2}$–$5P_{3/2}$–$5D_{5/2}$ transition of $^{87}$Rb. The excitation due to this two-photon absorption was collectively enhanced to generate photon pairs in the phase-matched direction from the Doppler-broadened atomic ensemble. Hence, highly bright photon-pair generation was achieved with a coincidence counting rate per input pump power of 64,600 cps/mW and a relatively long coherence time of 1.87 ns, under 48-mW coupling power. Under the conditions of 10-mW coupling power, the non-classical correlation between the photon pairs has been demonstrated by the strong violation of Cauchy-Schwarz inequality by a factor $R = 2370 \pm 150$. We confirmed that the photon pair generated in the Doppler-broadened atomic medium has a temporal non-classical quantum correlation. Furthermore, to verify the anti-bunching feature of the heralded single photon, we investigated the photon statistics for the heralded single photons by using the conditional HBT experiment and estimated the normalized second-order correlation function $g_C^{(2)}(0)$ of the heralded single photon as $g_C^{(2)}(0) = 0.037(3)$ with 0.2-mW coupling power. We believe that our photon-pair source may be applied to studies of quantum optics including quantum-entanglement swapping between completely autonomous sources.